\documentclass[twocolumn,pra,aps,superscriptaddress,showpacs]{revtex4-1}

\usepackage{lineno}
  \usepackage{mathptmx}
\usepackage{subfigure}
\usepackage{dcolumn}
\usepackage{amsmath,amssymb}
\usepackage{bm}
\usepackage{color}
\usepackage{overpic}
\usepackage{latexsym}
\usepackage{epstopdf}
\usepackage{color}
\usepackage[english]{babel}
\usepackage{latexsym}
\usepackage{stmaryrd}

\usepackage{psfrag,graphicx}
\usepackage{epsf}
\usepackage{subfigure}
\usepackage{amsmath}
\usepackage{amssymb}
\usepackage{amsfonts}
\usepackage{bm}
\usepackage{natbib}
\usepackage{epstopdf}
\usepackage{appendix}

\definecolor{mygrey}{gray}{0.35}
\definecolor{myblue}{rgb}{0.2,0.2,0.8}
\definecolor{myzard}{cmyk}{0,0,0.05,0}
\definecolor{mywhite}{rgb}{1,1,1}
\definecolor{myred}{rgb}{1,0.,0.3}

\usepackage[colorlinks=true,citecolor=myblue,linkcolor=myred]{hyperref}
\def\be{\begin{equation}}
\def\ee{\end{equation}}
\def\ba{\begin{align}}
\def\enda{\end{align}}
\def\bi{\begin{itemize}}
\def\ei{\end{itemize}}

 \def\ee{\mathord{\rm e}}

 \def\ee{\mathord{\rm e}}

\renewcommand{\ee}{{\rm e}}

\def\beq{\begin{equation}}
\def\beq{\begin{equation}}
\def\eeq{\end{equation}}

\begin{document}


\title[Short Title]{Quantum dynamics of bio-molecular systems in noisy environments}

\author{M.B. Plenio}
\affiliation{Institut f\"{u}r Theoretische Physik, Albert-Einstein Allee 11, Universit\"{a}t Ulm, 89069 Ulm, Germany}
\author{S.F. Huelga}
\affiliation{Institut f\"{u}r Theoretische Physik, Albert-Einstein Allee 11, Universit\"{a}t Ulm, 89069 Ulm, Germany}

\pacs{ }

\begin{abstract}
We discuss three different aspects of the quantum dynamics of bio-molecular systems
and more generally complex networks in the presence of strongly coupled environments.
Firstly, we make a case for the systematic study of fundamental structural elements underlying the quantum dynamics of these systems, identify such elements and explore
the resulting interplay of quantum dynamics and environmental decoherence. Secondly,
we critically examine some existing approaches to the numerical description of system-environment interaction in the non-perturbative regime and present a promising
new method that can overcome some limitations of existing methods. Thirdly, we
present an approach towards deciding and quantifying the non-classicality of the
action of the environment and the observed system-dynamics. We stress the relevance
of these tools for strengthening the interplay between theoretical and experimental
research in this field.
\end{abstract}

\maketitle


\section{Introduction}

At the fundamental level Nature is quantum mechanical and it is therefore not
surprising that quantum coherence and entanglement exist in well isolated multi-component
systems, such as the electrons involved in chemical bonds. One may argue, however, that in
such situations, when the system is static, coherence and entanglement can always
be made to vanish in a suitably chosen basis unless there is a natural (spatial) separation
into distinguished subsystems.  Hence, to observe non-trivial quantum effects we
need to force the system out of its eigenbasis, through intervention from the outside, and thus probe quantum coherences
between eigenstates as well as the quantum coherence properties of their consequent
dynamics. It is therefore the dynamics generated by the external perturbation
that results in relevant quantum phenomena and provides the means to probe for
interesting quantum properties.
The external perturbation may either be coherent and controlled, as it would be
favoured by an experimentalist aiming to interrogate the system, or it may be provided by an
interaction with an unobserved environment. The latter will always be present,
as it is never possible to completely isolate a physical system from its environment.
In fact, in bio-molecular systems the interaction with uncontrolled environments,
for example in the form of thermal fluctuations, is an important driving force of
its dynamics and hence its functionality. It may therefore be expected that the
interplay between the internal coherent quantum dynamics of the system and the
unavoidable presence of noise introduced by the environment has been optimized by
Nature and may be significant for its function. In fact, despite the tendency
for environmental noise to destroy coherence through the introduction of classical
randomness, it is also well known to have the capacity to lead to the generation
of coherence and entanglement and be instrumental for its persistence in the steady
state \cite{PlenioH02,HartmannDB06}.
It is then an interesting question to determine up to what extent are quantum dynamics, coherence and
entanglement present in composite quantum systems in contact with environments,
such as bio-molecular systems and, importantly, what role, if any, genuine quantum
traits may play for its functionality. These questions become particularly challenging
in bio-molecular systems where the strength of the system-environment interaction is
often comparable to the intra-system coherent coupling strengths. This non-perturbative
regime is most challenging, as it sits uncomfortably between the two standard limiting
cases in the system-environment interaction that admit relatively straightforward treatments.
Hence novel theoretical methods will need to be developed, a task that may benefit
from the increased understanding of quantum dynamics of composite quantum systems
that has emerged in recent years in quantum information science.


A variety of challenges present themselves. Firstly, composite quantum systems or
quantum networks in noisy environments will exhibit a wealth of dynamical features
whose complexity will increase very rapidly, perhaps even exponentially, with the
number of components thanks to the concomitant growth of the size of the state space.
One approach may simply consist in the development of increasingly detailed models
that reproduce the dynamics with the help of a sufficient number of free parameters.
While this may be suitable for smaller systems, it rapidly becomes neither workable
nor fruitful for larger ones - a situation not dissimilar from that encountered in
entanglement theory in quantum information science \cite{PlenioV07}. However, knowing
all the details, i.e. knowing all the parameters determining the dynamics, does not
necessarily lead to greater understanding. More might be learnt if one distils
key elements that govern the dynamics and that are broadly applicable across a large
number of systems. In the context of noise assisted transport, steps towards this goal
have been taken and will be outlined briefly.

Secondly, even if it is possible to deduce some of these key principles on the
basis of simplified models for the system environment interaction, it will become
necessary to put these principles to the test in increasingly realistic models of
the system-environment interactions and, crucially, in  actual experiments. This challenges
current methodology which has been derived largely for the limiting case of weak
or strong system-environment coupling. Hence new tools need to be developed for
the intermediate regime admitting increasingly precise descriptions together with,
ideally rigorous, error bounds to enable the reliable benchmarking with respect
to each other. In this note we will briefly discuss existing methods, pointing out
strengths and weaknesses and outline a novel approach that is bringing together methods
from mathematics, condensed matter physics and quantum information and holds the
promise of overcoming some of the limitations of existing methods.

Thirdly, the experimental and theoretical study of the dynamics of multi-component
systems in the presence of noisy environments raises the question of assessing at
what point the system and its dynamics may be considered quantum coherent and at
what point the system dynamics has become classical, in a quantitative fashion.
The very same questions may be raised concerning the action of the
environment acting on the system. In fact, judging reliably when it is really
necessary to model the environment quantum mechanically and providing a classification
of those physical quantities for which an accurate specification can be achieved by modelling the environment in terms of a fluctuating, possibly non-Markovian, classical field could lead to considerable
simplifications in the description of quantum dynamics in bio-molecular systems.
Furthermore, robust and well-founded methods for the quantification of the quantum
character of bio-molecular dynamics would provide conceptual tools to deliver
"extraordinary evidence for extraordinary claims" of the existence of quantum
coherence as required in \cite{Wolynes09}. Quantum information theory has longed
grappled with the quantification of quantum resources and we
will discuss these questions briefly bringing to bear methods from this discipline.

\section{Structural elements of noise assisted quantum dynamics}
Exciton energy transfer (EET) has
been actively studied within the chemical physics community for decades \cite{Forster48,Redfield65,GroverS71,Pearlstein72,HakenS73,KrenkeK74}, yet
the topic remains timely \cite{Scholes03,AdolphsR06,ChengF09}, mainly due
to the development of new spectroscopic techniques in the femptosecond time scale.
Ultrafast nonlinear spectroscopy has been used recently to probe energy transfer
dynamics in the Fenna-Matthew-Olson (FMO) and other photosynthetic aggregates \cite{BrixnerSV+05,EngelCR+07,LeeCF07,PanitchayangkoonHF+10,MercerEK+09}. The FMO
complex is an example of a pigment-protein complex (PPC), a network through which electronic
excitations on individual pigments can migrate via excitonic couplings. These experiments
have provided evidence for the existence of significant quantum coherences between
multiple pigments through the presence of wave-like beating between excitons which
has been observed to persist on timescales $> 550$fs, a significant fraction of the
typical transport time in FMO. As a result, there is now considerable interest in exploring
the possibility of assigning a functional role to quantum coherence in the remarkably efficient
excitation energy transfer in FMO and other PPCs.  Theoretical
investigations of the role of pure dephasing noise in EET have found that this noise
mechanism has the ability to enhance both the rate and yield of EET when compared to
perfectly quantum coherent evolution \cite{MohseniRL+08,PlenioH08}. In the exciton
basis normally used in previous studies \cite{ChoVB+05}, dephasing-assisted transport
(DAT) is considered to be resulting from noise-induced transitions between exciton
eigenstates states, which cause energetic relaxation towards the reaction centre.
This approach suffices to suggest the existence of EET as a noise-assisted processes
but it is less transparent in showing how DAT actually works, when it does not exist or is inefficient
and how it might be controlled or exploited in artificial nanostructures. Here we
would like to provide a fresh look at the problem by aiming at clearly identifying the
mechanisms that are underlying noise assisted transport
\cite{MohseniRL+08,PlenioH08,OlayaCastroLO+08,CarusoCD+09,ChinDC+10,RebentrostMK+09,CarusoCD+10,IshizakiCS+10}.
Identifying and spelling out these mechanisms clearly does provide additional
value even if the individual contributions might have been known in some context.
Such classifications pave the way towards tailored experimental tests of individual
contributions \cite{CarusoMC+10} while the clear understanding of these fundamental mechanisms
will allow for the optimization of network architectures in order to utilize quantum coherence
and noise optimally in artificial nanostructures. Here we briefly outline the
developments of \cite{PlenioH08,CarusoCD+09,ChinDC+10,CarusoCD+10}.

\subsection{Description of elements}
\begin{figure}[hbt]
\centerline{\includegraphics[width=7cm]{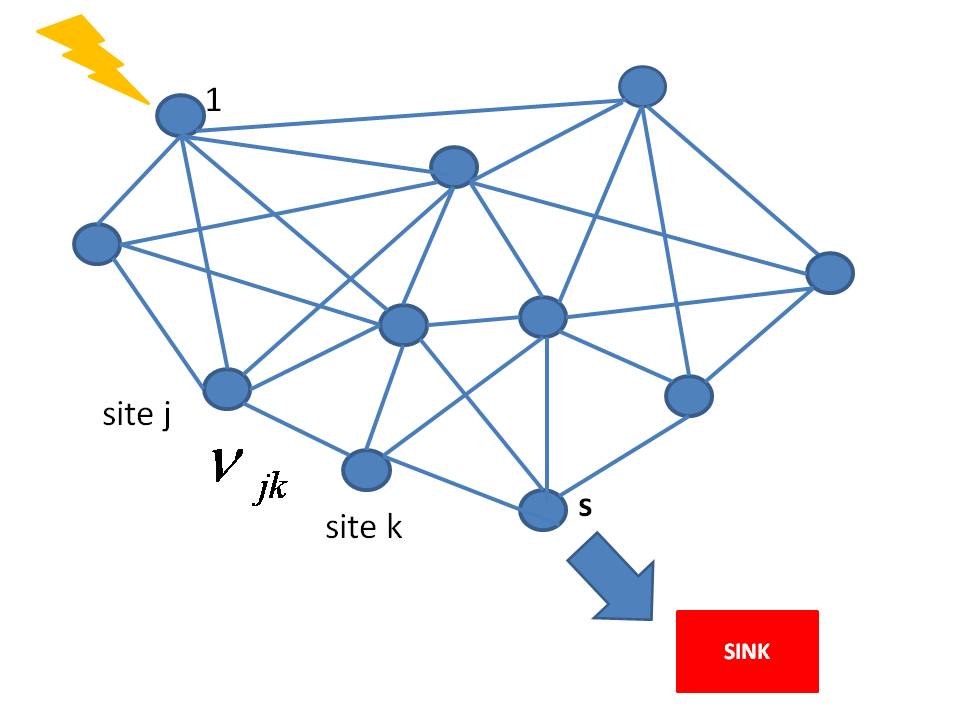}}
\caption{Quantum network of arbitrary geometry.  Network sites are modeled as
two level systems and excitations can propagate across the network via coherent
exchange interactions.  Select sites labelled 1 and s receive an initial excitation
and are irreversibly coupled to a sink respectively.  Quantum effects may significantly
inhibit transport, so that in the limit of large number of sites N, no population is
transferred to the sink.  Environmental noise in the form of pure dephasing can
significantly alter this pattern and assist efficient population transfer}
\label{fig1}
\end{figure}

The fact that noise can assist transport is well known in certain contexts. For
instance, incoherent hopping between neighbouring molecules can be facilitated
by the line broadening resulting from a variety of processes.  It is also well
established that interference effects in infinitely extended quantum networks
can preclude transport, as strikingly illustrated by the phenomenon of Anderson
localization in large lattices. We will summarize in this section recent results
showing that environmental noise can also assist the excitation transport across
finite networks and will identify the underlying basic mechanisms for DAT which
are also relevant for more general noise processes. With this background, we will
discuss a simple model of transport across the
Fenna-Matthew-Olsen (FMO)-complex, a bio-molecular wire present in types of
photosynthetic bacteria, and show that it's near unit transport efficiency can be
understood as the result of a dephasing assisted process. For clarity and simplicity
let us begin by considering a network of N sites such as the one illustrated in
Figure \ref{fig1} even though the considerations are general \cite{CarusoCD+09}. Each site
is modeled as a two level system corresponding to having either none or one excitation
at the corresponding site. Local sites couple to each other via a coherent exchange
interaction.  When each site couples to any other in the network with equal strength,
the network is said to be fully connected. The question we focus on is the following:
If site 1 is initially excited, what is the probability that at a subsequent given
time $t$ the excitation has been transferred to a sink to which a selected node $s$
is irreversibly coupled?
It can be shown that transport across this type of network is highly inefficient and
that the asymptotic sink population scales inversely with the number of sites so that
transport is essentially inhibited already for a moderate size network \cite{CarusoCD+09}. The fundamental reason behind this is the exact cancelation
of tunneling amplitudes of delocalized anti-symmetric states. It is remarkable
that thanks to the initialization of the excitation in a single site, the network
will generally have a high probability to evolve into a trapped state irrespective
of the detailed network structure. The coherent evolution
of the network therefore leads to the emergence of trapping states and precludes
excitation transfer to the sink, while purely incoherent hopping across the network
would eventually lead to complete transfer. It is now apparent that the transport
can become noise-assisted if a type of noise is provided that is able to disturb
the formation of invariant subspaces, which are decoupled from the sink. This is precisely
what local dephasing, and other forms of noise, can do. As opposed to relaxation,
dephasing noise leaves the site population unchanged but destroys the phase coherence
of superposition states. When subject to dephasing noise, cancelation of tunneling
amplitudes is no longer exact and the condition for excitation trapping is removed. Dephasing noise
is therefore able to open up transport paths that interference effects had rendered
forbidden and facilitate the transfer to the sink. The transfer efficiency becomes a
non-monotonic function of the local dephasing rates \cite{MohseniRL+08,PlenioH08,OlayaCastroLO+08,CarusoCD+09,ChinDC+10,RebentrostMK+09} so that an optimal
noise strength can be found for which transport is most efficient and at which quantum
coherence is partially but not completely suppressed \cite{CarusoCD+10}. Higher levels of noise lead to
the effective localization of the excitation and therefore transport is again compromised.
It should be noted that the decohering effect of strong noise may also be used
constructively in blocking undesirable paths for propagation in the system by
selectively applying noise to specific sites.

There are a number of additional building blocks that contribute to the dynamics
of such a quantum network. Destructive interference may also be obstructed by the
presence of static disorder in the network as this leads to the development of
time dependent relative phases at a rate proportional to the energy difference
between sites. This in turn rotates the system out of trapped states into
propagating states and
hence aides transport. However, there are limits to the efficiency of this mechanism, as energy gaps that are larger
than the coupling matrix elements will lead to suppression of transport due to
energy conservation.
Here noise may again play a constructive role as fluctuations of energy levels
lead to line broadening and hence an enhanced overlap between lines. Alternatively,
one may view this process dynamically and realize that, thanks to the fluctuations,
energy levels sometimes become near degenerate and hence permit transport.
Excessive fluctuations however will lead to small time intervals in which levels
are near degenerate and hence transport will reduce again leading to a non-monotonic
behaviour of the excitation transfer as a function of the noise strength.
It should be noted that also quantum coherent interactions generally shift and
split energy levels. This can provide functional advantages as it may lead to a
broader absorption spectrum (e.g. in the LHII system) and may, as we will see
below, also move energy levels between which transfer is desirable closer to each other. On
the other hand, it unavoidably also shifts energy levels further apart and hence noise may again be needed to bridge those gaps.

\begin{figure}[hbt]
\centerline{\includegraphics[width=10cm]{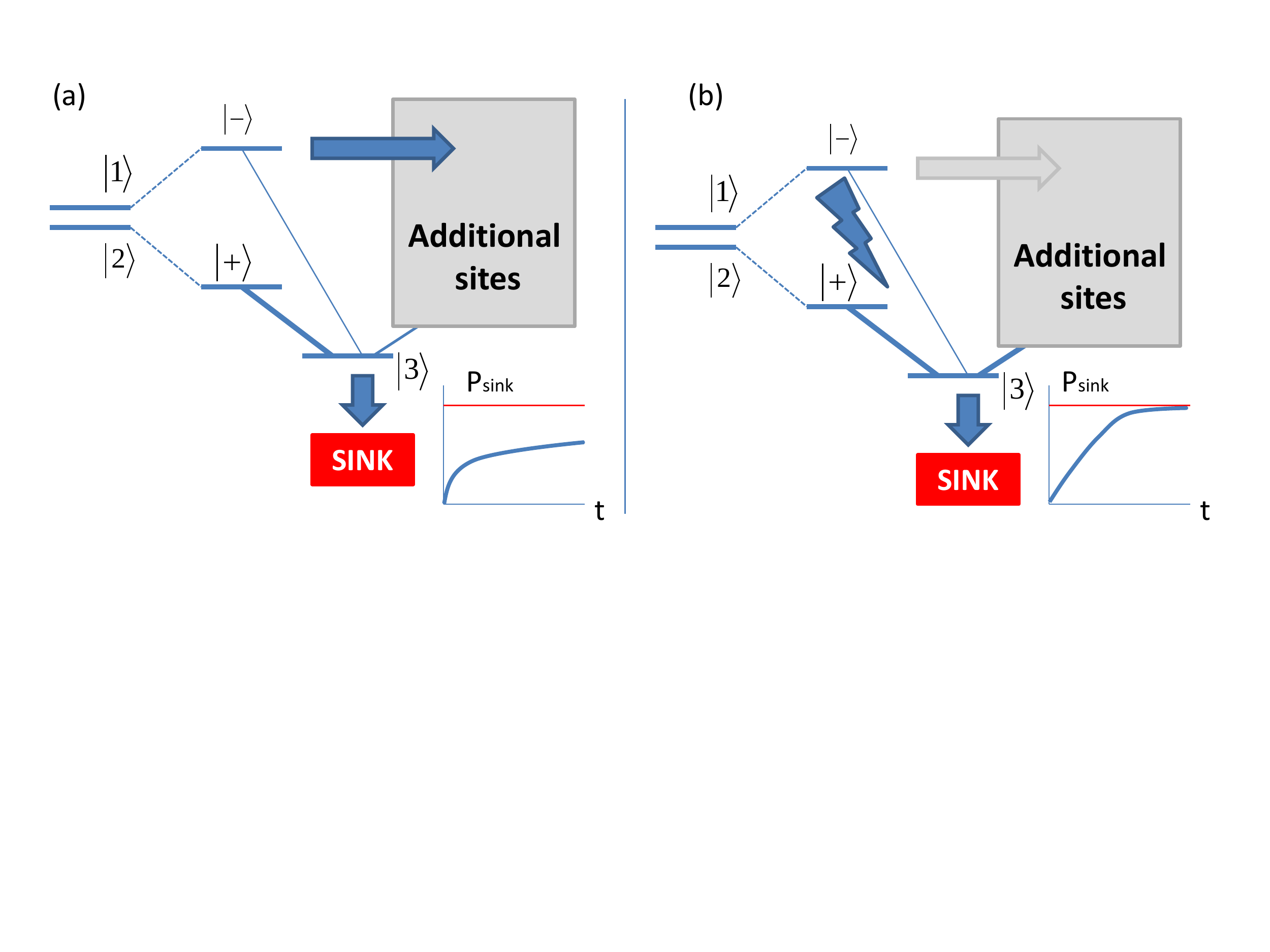}}
\vspace*{-3.cm}
\caption{Noise-assisted energy transfer across FMO can be qualitatively understood
by introducing a hybrid basis of local sites. While a purely unitary evolution yields
to inefficient transport, as discussed in the text, the presence of dephasing noise
eliminates inefficient transport paths while opening up new channels for excitation
transfer.  While the estimation for the transfer efficiency under coherent evolution
is well below 100$\%$, a simple noise model brings this number close to perfect efficiency
within the observed transfer time.}
\label{Fig2}
\end{figure}

We argue that these basic principles also apply for general networks. To illustrate how the
general principles outlined above may feature in a more realistic scenario, we
will consider briefly the example of excitation transfer across a monomer of
the FMO complex, which can be modeled as a $7$ sites network and for which detailed
information concerning the system Hamiltonian is available \cite{AdolphsR06}.
Given the strong coupling of sites $1$ and $2$ these levels are shifted and mixed
and the dynamics is conveniently described using an hybrid basis \cite{ChinDC+10}
for these two sites that we denote by $|\pm\rangle$. In this basis, the Hamiltonian
has the local site energies and coupling structure shown in Figure \ref{Fig2}, where
site $3$ is connected to a sink node that models the actual reaction center and
all the remaining sites have been packed in a block (labeled as "additional sites"
in Figure 2) that is uncoupled from the level $|+\rangle$.

When an excitation is injected in site 1, the coherent evolution leads to transfer
efficiency $P_{sink}$ below $60\%$, far from the ideal transfer, represented in
Figure \ref{Fig2} by the red horizontal line.  The coherent interaction between sites
$1$ and $2$ leads to level splitting and moves one of these energetically closer to
site $3$ while the other is farther removed. Furthermore, interference of transition
amplitudes from the states $|+\rangle$ and $|-\rangle$ will contribute to the dynamics
as explained above. The introduction of local dephasing noise can dramatically alter
this picture and lead to perfectly efficient transport to the sink, as illustrated in
the part (b) of the Figure. Coherent oscillations between level $|-\rangle$  and the rest of the
complex are now largely suppressed, while an incoherent transfer path between previously
decoupled levels $|\pm\rangle$ is now active, leading to a fast transfer of population to the sink.
Furthermore, destructive interference between propagation paths is reduced. Within this
model, it is possible to optimize the local dephasing rates so as to reproduce the
observed transfer rates in the correct time scale. More detailed discussions can be
found in the literature \cite{MohseniRL+08,PlenioH08,CarusoCD+09,ChinDC+10}.

\subsection{Remarks}
There are two aspects that we should stress in order to conclude this section. The
first one concerns the independence of the main results from the very specific details of the
noise model; while initial work involved simple dynamical evolutions of the Lindblad
form \cite{MohseniRL+08,PlenioH08}, subsequent studies refined the noise model to
also account for certain non-Markovian effects \cite{ChinDC+10}. Secondly, the range of noise
parameters leading to efficient transport is rather broad so that the interplay between
coherent and noisy processes does not require a fine tuning, but is robust to
variations in both the system Hamiltonian and the parameters characterizing the
system-environment coupling. These results challenge the traditionally held view
that noise tends to degrade the efficiency of quantum processes, and demonstrates
that controllable noise can even be considered as an additional engineering tool
for excitation transport \cite{MohseniRL+08,PlenioH08,CarusoCD+09,ChinDC+10,RebentrostMK+09,CarusoCD+10}.
Needless to say, in addition to these theoretical calculations it should be crucial
to provide a direct experimental test for the relevance of noise assisted transport
as well as the basic mechanisms that underly the dynamics of the FMO complex. Here
techniques from optimal control for state preparation may become crucial to realize
recent proposals towards this goal \cite{CarusoMC+10}.

\section{Numerical simulation of noisy quantum dynamics}
The previous section has demonstrated that the interaction between system and
environment can be of crucial importance for both dynamics and functionality
of the system. Furthermore, the optimal operating regime, at least with respect
to transport, is found to be one in which the strength of the system-environment
interaction, characterized by the so-called reorganization energy, is comparable to the
intra-system coupling rates. This regime is quite different to the two extreme
cases in the system environment interaction that are usually studied and which
admit perturbative treatments. On the one hand, very strong coupling
leads to a dynamics that is well approximated by classical rate equation models.
Weak system environment coupling on the other hand permits a perturbative treatment
in which environmental dephasing and relaxation can be treated with Lindblad or
Bloch-Redfield master equations which are both based on the assumptions of weak
system bath coupling and the Markov approximation. The intermediate regime, however,
appears to be of particular relevance to bio-molecular systems but at the same
time poses greater challenges as it is intrinsically non-perturbative. Here
perturbative expansions that assume environments with correlation times that
are much smaller than the system dynamics are not well justified. Indeed, in
the limit of slow bath dynamics, perturbative treatments of the system environment
coupling cannot be used even if the system-bath coupling is intrinsically weak.
Hence, in the light of its importance the development of methods for the exact
and flexible description of system-environment interaction in this regime is
well motivated.
In fact, the development of such methods has gathered pace and recently, and a variety
of steps have been taken towards the development of fully non-perturbative and
non-markovian approaches \cite{IshizakiF09a,PriorCH+10,ChinRH+10,RodenEWS09,NalbachT10}. The relative merits of the different approaches have
not been compared and contrasted yet, nor have the regimes in which they may
be optimal been identified. This would be a timely issue as it appears likely
that different approaches will be best suited for different physical settings
and different quality measures and requirements.

In the following we discuss three of the approaches that are being pursued in the
literature and suggest strengths and weaknesses of these methods.

\subsection{The hierarchy method}
Recently, the numerical hierarchy technique
\cite{IshizakiF09a,IshizakiF09b,IshizakiT05,DijkstraT10}, which has a longstanding
history \cite{TanimuraHK77,TanimuraK89,Tanimura06}, has received renewed attention
in the context of EET across pigment-protein complexes. This approach is non-perturbative and is capable of interpolating, for example, between the Bloch-Redfield and the F{\"o}rster regimes \cite{IshizakiF09a,IshizakiF09b}. It
derives a hierarchy of equations in which the reduced density operator
of the system couples to auxiliary operators which in principle allow for the
simulation of complex environments. The depth of this hierarchy and the
structure of its coefficients depend on the correlation time of the bath and its
spectral density. This approach appears sufficiently flexible to take account of
spatial correlations in the noise as well. For specific choices of the bath spectral
density, namely the Brownian harmonic oscillator, and some mild approximations, one obtains
a relatively simple structure of the hierarchy. In this case, the temporal bath
correlations decay exponentially so that the hierarchy can be terminated early
with small error. An estimate suggests that in this case the set of operators in the
hierarchy will scale at least with $\tau^k$ where $\tau$ is the correlation time
and $k$ is the number of sites in the system to be studied (see \cite{IshizakiCS+10}
for a more detailed discussion). More complex spectral densities will lead to
considerably more demanding evaluations of the coefficients of the hierarchy. A
non-exponential decay of the temporal bath correlations also leads one to
estimate an exponential growth of the number of operators in the hierarchy. It is
possible that specific numerical simulations turn out to be more efficient than
those estimates suggest but there is no certificate that provides error bounds.
In fact, it is a challenge to provide rigorous bounds of
the errors introduced by the various approximation steps that are involved in the
numerical hierarchy technique. Hence one has to test for convergence empirically by
increasing the depth of the hierarchy until the result do not change significantly
anymore. It should be noted, that the hierarchy method has been applied successfully
to a dimer \cite{IshizakiF09a,IshizakiF09b} as well as the seven-site FMO complex
\cite{IshizakiF09c} with a Brownian harmonic oscillator spectral density of the
environment.

\subsection{Path integral methods}
A variety of other approaches to study transport processes and in consequence also
transport in noisy environments have been developed in condensed matter physics.
Recently some of these have been applied to dimer systems with the aim of understanding
better the evolution of quantum coherences in structured environments. These methods
represent various approaches for solving numerically the formal path integral solution
of the time evolution. These include the quasi-adiabatic path integral approach \cite{MakarovM94,MakriM95a,MakriM95b,NalbachT10,ThorwartER+09} and the iterative
summation of real-time path integrals \cite{WeissET+08} to name just two approaches.
These procedures are expected to give good results in the high temperature limit and
hence short correlation time of the environments. With decreasing temperatures the
computational effort is growing rapidly and the $T=0$ limit cannot be reached, while
temperatures not too far below the typical system frequencies appear to be accessible \cite{ThorwartRH00}.
For highly structured environments in which for example both narrow and broad features
are combined these methods find challenges. In this case, sharply peaked modes can be
added to the system and their damping is treated in the bath \cite{ThorwartPG04} but
such an approach will be problematic for the treatment of quantum networks when addition
of modes make the network itself too high-dimensional for numerical treatment.
Path integral methods have been applied mostly to dimers (see e.g.\cite{ThorwartER+09})
and their scaling to larger systems, just as for the transformation approach to be discussed below, remains to
be demonstrated.

\subsection{The transformation technique}
An alternative approach that has the potential to address some of the challenges
that we have raised above has been proposed recently. It treats the spin-boson
model by a combination of analytical transformation techniques (building and
improving upon earlier related work \cite{BullaTV03,BullaLT+05}) to make it amenable
to simulation methods from condensed matter physics and quantum information theory
\cite{PriorCH+10,ChinRH+10}.
In the spin-boson model, each two-level system couples linearly to the coordinates
of a separate environment consisting of a continuum of harmonic oscillators presented
schematically in Figure 3.a for the situation of a dimer. This model represents a
suitable starting point for the description of excitation transport in pigment-protein
complexes and is capable of encompassing a wide variety of dynamic behavior including
non-Markovian dynamics as the vast majority of environmental modes in a pigment-protein
complex are to a very good approximation harmonic \cite{HaywardG95}. Numerical treatments
of the spin-boson model can be loosely divided into methods in which the environmental
degrees of freedom are eliminated in order to derive effective equations of motion for
the two-level system, and those in which the complete many-body dynamics of the
two-level system and the environment are considered.
\begin{figure}[hbt]
\vspace*{-0.3cm}
\centerline{\includegraphics[width=9.cm]{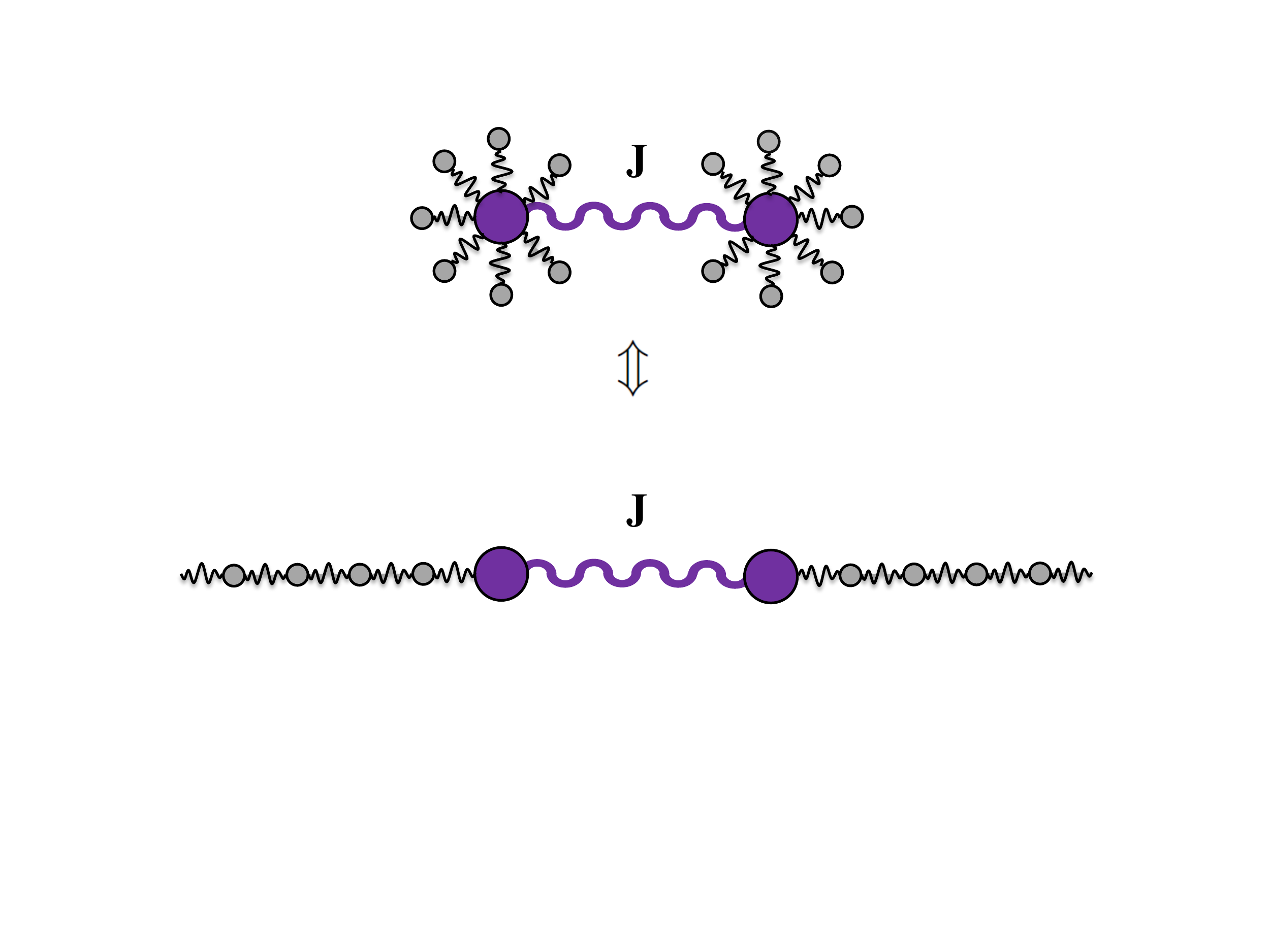}}
\vspace*{-2.cm}
\caption{(a) Atop, in the standard representation of the spin-boson model considers central systems (grey spheres) coupled individually to a bath of bosonic modes representing harmonic oscillators. (b) Below, after an exact unitary transformation, solely affecting the bath degrees of freedom, the same number of bosonic modes are now arranged in a linear chain with nearest neighbour  interactions only. The latter setting admits straightforward numerical description of the full dynamics with the time dependent density renormalization group method.}
\end{figure}
The transformation approach treats the full system environment dynamics. In a
first step the configuration depicted in Figure 3.(a) is transformed into the
linear configuration shown in Figure 3.(b) where each system site couples to a
linear chain of bosonic modes with a nearest neighbor hopping Hamiltonian. This
transformation, which can be determined analytically for a wide variety of
spectral densities, or can otherwise be obtained in a numerically stable and
efficient way, acts only on the environment degrees of freedom. As a consequence,
the dynamics of the system, e.g. the pigment protein complex is unchanged. The
dynamics of the resulting system + environment in Figure 3.(b) now has the
appropriate structure (linear chain with nearest neighbor interactions) so
that it can now be efficiently integrated numerically employing the time-dependent
density matrix renormalization group algorithm \cite{DaleyKS+04}. The complexity
scales linearly in the size of the environments.

This approach has several
advantages. Firstly, the transformation approach works for arbitrary and possibly
highly structured spectral densities of the environment without additional
computational overhead. In particular, it is capable of including narrow features
in general backgrounds and spectral densities leading to power law decay of
the bath correlation function \cite{PriorCH+10} without sacrificing its
efficiency. Secondly, the time-dependent density matrix renormalization group
algorithm and hence the transformation approach can be made arbitrarily precise
(at the cost of what is usually a polynomial increase in computation time) and,
crucially, delivers in each time step and hence also for the total evolution, an
upper bound on the error committed. Hence the simulation delivers rigorous error
bars from a single run. Finally, the transformation approach provides the full
information about system and environment and their dynamics leading to a better
understanding, for example, of the irreversibility of the system-environment
interaction \cite{ChinRH+10} but also opens the possibility of studying system-environment
interactions in which the bath is prepared in non-trivial initial states or
in which knowledge about the state of the environment should be extracted.

While the transformation method has been demonstrated for a dimer in contact
with zero temperature environments, \cite{PriorCH+10,ChinRH+10} it is not
restricted to this setting. It can be generalized to multi-site systems and
finite temperatures at the expense of a moderate increase in computation time
\cite{RosenbachPC+10}. It should be noted however, that at present the inclusion
of spatial quantum correlations between environments is challenging and restricted to
simple cases. This may be a drawback for certain applications in which spatial
correlations play a significant role but this does not appear to be the case
in photosynthetic complexes like FMO. It should also be noted that the transformation
approach immediately yields systematic approximation techniques (see also
\cite{HughesCB09a,HughesCB09b}) by cutting short the chains that are obtained
in the exact transformation and coupling the last site to a Markovian bath
\cite{ChinRH+10}.

\subsection{Remarks}
We have briefly outlined and discussed possible approaches for the numerical
and analytical study of non-perturbative system-environment interactions with
a view to applying these methods to pigment-protein complexes. Each of these
methods has been applied separately and certain strengths and potential weaknesses
seem to emerge. However, as already mentioned before, a comprehensive comparison of all of these approaches has
not been undertaken yet. Such a comparison would be highly desirable as it would
allow us to ascertain the validity of the various approaches and also permit
us to identify the most suitable methods for each possible experimental setting
and for various success parameters. This would allow us to study a wide variety
of questions of applied and fundamental nature. In particular, we may be able to
consider the question to what extent the action of an environment is in fact
quantum mechanical and what degree of approximation we can achieve with classical
environment models. This and related questions will be discussed in the final
section.

\section{Assessing classical and quantum dynamics}
Recent experiments are providing supporting evidence for the existence of quantum
coherence and quantum coherent dynamics in a variety of bio-molecular systems despite
the presence of strong coupling to the environment. The inferences concerning the quantum
character drawn from these observations are however qualitative; the existence of
oscillating features or off-diagonal elements in 2-D spectra serve as evidence that
some quantum coherence is present or some quantum dynamics is taking place but
does not make these statements quantitative. To pursue these questions in a more quantitative fashion it will be
necessary to transfer tools from quantum information science that allow us to infer
from the experimental data quantitative upper and especially lower bounds on the
quantum mechanical properties of the system, ideally without invoking any experimentally
untested assumptions. This might be done by studying entanglement measures \cite{CarusoCD+09,CarusoCD+10,SarovarIF+10,FassioliO10} but this approach suffers
some drawbacks. Firstly, entanglement measures are often difficult to measure
directly in actual experiments. Secondly, entanglement measures have been defined in quantum
information science in a specific context, namely to study quantum information
tasks in which we have a natural separation into subsystem (usually defined by
distance) \cite{PlenioV07}. It is not clear that this approach is directly relevant
to the study of quantum effects in biology.
Here we would like to indicate briefly two aspects of this problem and possible ways
forward that we are pursuing. Experimentally, we tend to be limited to snapshots of
the dynamics at different times. On the basis of these observations we would like to
quantitatively assess (a) whether the action of the environment can be modeled classically
and if not, to what extent it deviates from being classical and (b) to what
degree the dynamics of the system alone is quantum mechanical. Questions of this type
have been posed before in quantum information science and we will be making use of the techniques developed there (see \cite{RivasAP+10} for details).

\subsection{Classical Environments}

Ultimately, all dynamics needs to be described in the formalism of
quantum mechanics and to make progress we define what
is meant by a classical action of an environment in this framework. Intuitively,
an environment should be termed classical if we can model it by a device that
generates random numbers $i$ with probability $p_i$ and on the basis of these
then applies a unitary action $U_i$ to the system. Correlations between system
and environment then remain classical at all times. Formally, if as observers
we are ignorant of the value of $i$, such an action would result in the map of
the system density operator
$\rho\mapsto\Phi(\rho)=\sum_i p_i U_i \rho U_i^{\dagger}$. That is, the evolution
can be expressed as a convex sum of random unitaries (a bistochastic map). Hence,
if we measure experimentally for a certain time $t$ the map $\Phi_t$ we can then
determine whether this map can be written in the form $\sum_i p_i U_i\rho U_i^{\dagger}$
employing numerical methods \cite{AudenaertS07}. If this is not possible,
the action of the environment is  non-classical. Remarkably, in the case that $\rho$
represents a two-dimensional Hilbert space and the evolution is such that the identity
is preserved (unitality condition), the resulting dynamics can always be expressed as a bistochastic map, while for any Hilbert space dimension larger than $2$, exceptions exist. This does not however imply that the environment action on the system is classical if we perform more demanding experimental tests. After all, if we attempt to simulate the action of the environment, we need to do so not only mapping initial state into the final state but need to progress along a time sequence $t_0,t_1,\ldots,t_{fin}$. While at the initial time $t_0$ the system might be decorrelated from the environment, it will generally not be so at later times. Then it is not evident anymore that the map $\Phi_{st}$ taking the system from time $s$ to time $t$ will also be classical, even if the maps $\Phi_t$ and $\Phi_s$ are classical separately. Hence, if we probe the system at times $t_0,t_1,\ldots,t_{fin}$ and determine the maps taking it from one time to the other and each of those maps is classical in the above sense then, for this time coarse graining, we would call the action of the environment on the system classical. This still leaves the question of the quantification of the degree of non-classicality of the environment action. As in the theory of entanglement, i.e. non-classical correlations, there may exist several possible quantifications. Taking an axiomatic approach one should define desirable properties of a measure of non-classicality of maps. The measure $C(\Phi)$ should vanish for maps of the form $\Phi(\rho) = \sum_i p_i U_i \rho U_i^{\dagger}$ which we have termed classical, it should also be such that for any $\Psi$ the measure does not increase under composition with a classical map, i.e. for any $C(\Phi\circ\Psi) < C(\Psi)$. In analogy with the theory of entanglement, one may then define a measure $C(\Psi)$ as the distance of the map $\Psi$ to the set of all classical maps \cite{VedralP98}. The distance between maps is not uniquely defined but a suitable choice might be the relative entropy distance between the Choi-Jamiolkowski states
\cite{PlenioV07} associated with these maps. Once such a measure is defined, it can also be used to obtain lower bounds on the non-classicality of maps by identifying the least non-classical maps compatible with the known parameters, quite in the spirit of related work in entanglement theory
\cite{AudenaertP06,EisertBA07,GuhneRW07}.

\subsection{Classical System Dynamics}
Needless to say, a classical action of the environment does not imply that
the system dynamics itself is classical. In fact, the most classical setting
is an environment that is decoupled from the system which allows the system
to evolve coherently. Questions of this type are of relevance in quantum
information where for example the quantum character of a memory needs to be
assessed on the basis of a number of preparations and subsequent measurements
\cite{OwariPP+08,JensenWK+10} and the question whether a dynamics exhibits
quantum coherence or not may then be approached with similar tools. Let us
start again by probing the dynamics of the system at one particular moment
in time, t. We would like to designate a map classical if it can be replicated
by a device that measures the system state initially and then produces a new
state from the measured data only. This so-called measure and prepare procedure
is of the form $\Phi(\rho) = \sum_i tr[\rho M_i]\sigma_i$ with positive linear
operators $M_i$ satisfying  $\sum_i M_i = 1$. It will not be able to perfectly
reproduce any possible map that can be realized in quantum mechanics as it
necessarily introduces some noise in the system due to the indistinguishability
of non-orthogonal quantum states. Again, the quantification of the non-classicality
of the map can be achieved by employing measures derived via an axiomatic approach.
The measure $C(\Phi)$ should vanish for maps of the form $\Phi(\rho) = \sum_i
tr[\rho M_i]\sigma_i$, it should also be such that for any $\Psi$ the measure
does not increase under composition with a map of the form $\Phi(\rho) = \sum_i
tr[\rho M_i]\sigma_i$,, i.e. for any $C(\Phi\circ\Psi) < C(\Psi)$ and the measure
is then defined as the distance between the map and the closest map of the form
$\Phi(\rho) = \sum_i tr[\rho M_i]\sigma_i$. Probing the system at multiple times
$t_0,t_1,\ldots,t_{fin}$ may be treated straightforwardly in this setting too.
It is useful to know that the determination of maps of this type is closely related
to problems of convex optimization whose methods can be made use of here \cite{RivasAP+10}.

\section{Conclusions}

Recent experimental results providing evidence of coherent behavior in EET across bio-molecular complexes have open up a most interesting path for multidisciplinary research.
Subsequent theoretical work has clarified the importance of the interplay between environmental noise and the underlying quantum dynamics in these systems.
However, the fundamental question of quantitatively linking quantum behavior and biological function remains open. In our view, developing novel techniques for optimally probing complex interacting quantum systems and efficiently processing the resulting experimental data will be crucial for finally discerning whether or not do quantum effects play a fundamental role in the dynamics of biological systems.

\section{Acknowledgements}
This work was supported by the EU Projects (CORNER, QAP and Q-ESSENCE), a
Marie-Curie Fellowship and an Alexander von Humboldt-Professorship. The work
reviewed here was carried out with our colleagues, postdocs and
students K.M.R. Audenaert, F. Caruso, A. Chin, A. Datta, J. Prior, A. Rivas
and R. Rosenbach.

\end{document}